\documentclass[
 aip,
 apl,       
 amsmath,amssymb,
 reprint,   
]{revtex4-1}

\usepackage{graphicx}
\usepackage{dcolumn}
\usepackage{bm}
\usepackage[utf8]{inputenc}
\usepackage[T1]{fontenc}
\usepackage{mathptmx}
\usepackage{etoolbox}
\usepackage{xcolor}
\makeatletter
\def\@email#1#2{%
 \endgroup
 \patchcmd{\titleblock@produce}
  {\frontmatter@RRAPformat}
  {\frontmatter@RRAPformat{\produce@RRAP{*#1\href{mailto:#2}{#2}}}\frontmatter@RRAPformat}
  {}{}
}%
\makeatother
\begin{document}

\preprint{AIP/123-QED}

\title{Anomalous Hall transport in Mn$_{3}$Sn$_{0.5}$X$_{0.5}$C (X = Ge and Zn)}

\author{Sunil Gangwar}
\affiliation{ 
School of Physical Sciences, Indian Institute of Technology Mandi, Kamand, Mandi-175075 (H.P.) India}
\author{C. S. Yadav*}
 \email{shekhar@iitmandi.ac.in}
\affiliation{ 
School of Physical Sciences, Indian Institute of Technology Mandi, Kamand, Mandi-175075 (H.P.) India}
\affiliation{ 
Center for Quantum Science and Technologies, Indian Institute of Technology Mandi, Kamand, Mandi-175075 (H.P.) India}
\date{\today}

\begin{abstract}
 Mn-based antiperovskites that exhibit topological surface states show potential applications in spintronics, magnetoelectronics, and quantum devices owing to the interplay of magnetism and topological properties. In this family of compounds, Mn$_{3}$SnC has a concurrent ferromagnetic (FM)/antiferromagnetic(AFM) ground state below $T$ $\sim$ 285 K, along with the Berry curvature driven anomalous Hall effect (AHE). Here, we present AHE in Ge and Zn doped Mn$_{3}$SnC compounds: Mn$_{3}$Sn$_{0.5}$Ge$_{0.5}$C (MSGC) and  Mn$_{3}$Sn$_{0.5}$Zn$_{0.5}$C (MSZC). MSGC undergoes paramagnetic (PM) to a concurrent AFM/FM transition at $T_{C}$ $\sim$ 300 K, whereas MSZC exhibits PM to FM transition at $T_{C}$ $\sim$ 240 K, followed by FM to ferrimagnetic transition at $T_{N}$ $\sim$ 170 K. The electronic transport in these compounds is influenced by the electron-phonon and electron-magnon scatterings and exhibits anomalous Hall resistivity ($\rho^A_{xy}$). Our study suggests that AHE in these compounds arises due to skew scattering and intrinsic Berry curvature mechanisms, and Electron-phonon and electron-magnon scattering play an important role in skew scattering at  high temperature. The doping of Mn$_{3}$SnC with Ge and Zn atoms notably enhances the value of its anomalous Hall conductivity.
\end{abstract}

\maketitle

Antiperovskite compounds have garnered significant research interest in recent years because of their unique band structure and anomalous transport properties. Antiperovskites, also known as inverse-perovskites are recognized with general formula M$_{3}$XA (M: Mn, Ni, Fe, etc; X: Ga, Cu, Sn and Zn; A: N, C and B), have interchanged cation and anion positions within the unit cell for the perovskite structure \cite{wang2020antiperovskites}. Specifically, Mn-based nitride and carbide antiperovskites Mn$_{3}$XA (A = N, C) possess octahedra Mn$_{6}$N or Mn$_{6}$C structures, with six Mn atoms situated at the corners of the octahedra, that are susceptible to magnetic frustration \cite{dias2015effect,deng2023magnetic}. The direct exchange between Mn-Mn atoms contributes to ferromagnetism, while the superexchange interaction between Mn-C-Mn leads to antiferromagnetism \cite{dias2015effect,dias2017mechanism}. The competition between these exchange interactions gives rise to a variety of magnetic structures, including collinear antiferromagnetic (AFM), collinear ferromagnetic (FM), collinear ferrimagnetic (FIM), non-collinear magnetic, and non-coplanar magnetic spin configurations. Magnetic states of these compounds are highly sensitive to variations in pressure, composition, and grain size \cite{deng2023magnetic}. This coordination coupled with their magnetic response, gives unique properties such as giant negative thermal expansion \cite{kodama2010gradual}, large magnetocaloric effect \cite{tohei2004large}, giant magnetoresistance \cite{kamishima2000giant}, anomalous Hall effect (AHE) \cite{torres2022anomalous}, anomalous Nernst effect (ANE) \cite{zhou2020giant}, superconductivity \cite{hoffmann2022superconductivity} etc. 

The Mn$_{3}$XC (X = Sn, Zn, Ga) compounds have a cubic crystal structure with the Pm-3m space group. Mn$_{3}$ZnC exhibits FM transition at $T_{C}$ $\sim$ 420 K, and FIM transition at $T_{N}$ $\sim$ 220 K \cite{teicher2019weyl}. On the other hand, both FM and AFM states coexist below $T$ $\sim$ 285 K in Mn$_{3}$SnC \cite{gangwar2024magneto}. In the concurrent AFM/FM state of Mn$_{3}$SnC, one Mn atom is aligned ferromagnetically along the [001] direction of the crystal lattice, with a magnetic moment of $\sim$ 0.7 $\mu_{B}$, while the other two Mn atoms are arranged antiferromagnetically in a square configuration, with a magnetic moment of $\sim$ 2.3 $\mu_{B}$ \cite{dias2015effect}. 

In the ordinary Hall effect, a transverse voltage appears as charge carriers experience Lorentz force in a magnetic field. However, in FM or topological materials, an extra transverse voltage is also generated due to intrinsic magnetic field, leading to AHE. AHE can be understood to arise from three mechanisms; (i) intrinsic Berry curvature, (ii) skew scattering and, (iii) side jump mechanisms. In intrinsic mechanism, charge carriers acquire anomalous velocity in an electric field due to the Berry curvature, which depends on the spin–orbit coupled electronic band structure. The sum of the velocities of all occupied Bloch electrons results in the anomalous Hall conductivity (AHC) \cite{nagaosa2010anomalous,yue2017towards}. Skew scattering and the side-jump mechanisms arise from the asymmetric scattering of charge carriers in the presence of spin-orbit coupling (SOC), and are categorized as extrinsic mechanism. Skew scattering depends on various factors such as; defects, impurities, electron-phonon scattering, spin-dependent scattering, and spin clusters \cite{yu2025large,yue2017towards}. AHE in Mn${_3}$SnC is primarily attributed to intrinsic Berry curvature at high temperatures, with extrinsic contributions at lower temperatures. Though, the value of $\rho^A_{xy}$ in Mn$_{3}$SnC is low, ($\sim$ 0.3 $\mu\Omega$-cm), the non-stoichiometry of carbon may give rise to enhanced extrinsic contribution in Mn$_{3}$SnC$_{0.8}$ ($\sim$ 0.6 $\mu\Omega$-cm) \cite{huang2022magnetoresistance}. Further, the observation of Berry curvature contribution in the Mn$_{3}$SnC reflects the topological aspect of the material. For Mn$_{3}$ZnC, Teicher $et$ $al$. reported it to show topological properties of nodal line semimetal (NLSM) with drum-head surface states and finite Weyl nodes \cite{teicher2019weyl}. The low temperature noncollinear magnetic ground state, topological state and the role of carbon stoichiometry are quintessential for understanding the AHE in Mn$_{3}$SnC and other antiperovskite materials.

Here, we present anomalous Hall transport studies of Mn$_{3}$Sn$_{0.5}$Ge$_{0.5}$C (MSGC) and Mn$_{3}$Sn$_{0.5}$Zn$_{0.5}$C (MSZC) compounds. We prepared polycrystalline samples of MSGC and MSZC using the solid-state reaction method. The preparation condition, structural and stoichiometric details are given in supplementary information (See Fig S1, S2 and S3). Substitution of Ge and Zn influences the magnetic phase transition, increases the AHC, and enhances the electron-phonon contribution to the longitudinal and Hall resistivity of Mn$_{3}$SnC. The value of AHC increases to 26 $\Omega^{-1}$cm$^{-1}$ in MSGC and 200 $\Omega^{-1}$cm$^{-1}$ in MSZC, that are significantly higher than $\sim$ 0.78 $\Omega^{-1}$cm$^{-1}$ for Mn$_{3}$SnC. 

\begin{figure}
	\begin{center}
		\includegraphics[width=8.0cm]{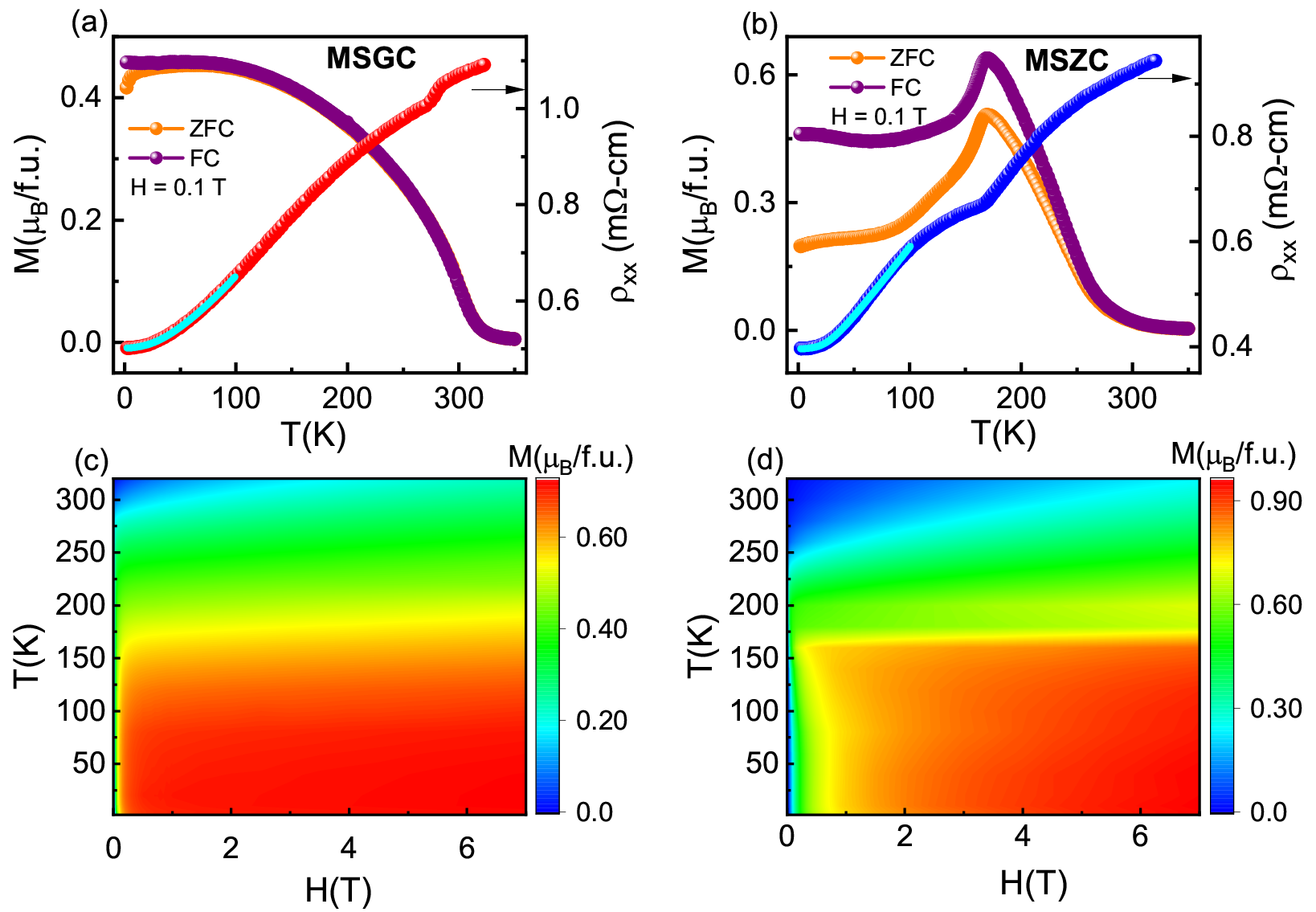}
		\caption{\label{Fig1}(a) and (b) Temperature-dependent magnetization $M$ and longitudinal resistivity $\rho_{xy}$ for MSGC and MSZC. Cyan line shows the fit line. (c) and (d) Magnetization vs. Field at different temperatures for MSGC and MSZC. }
	\end{center}
\end{figure}

 Figure 1(a) and (b) show the temperature variation of magnetization ($M(T)$) at H = 0.1 T field in zero-field cooling and field cooling protocols for MSGC and MSZC, respectively. For MSGC, a concurrent AFM/FM transition is observed at $T_{C}$ $\sim$ 300 K. Substitution of Ge at Sn sites raises the magnetic transition temperature by $\sim$ 15 K compared to Mn$_{3}$SnC \cite{huang2022magnetoresistance,gangwar2024magneto}. For MSZC, $M(T)$ shows a peak at $T_{N}$ $\sim$ 170 K, corresponding to FIM to FM transition, followed by a FM to PM transition at $T_{C}$ $\sim$ 240 K. Thus, Zn doping strengthens the AFM ordering and significantly lowers the $T_{C}$ of Mn$_{3}$SnC. Alternatively, it can stated that both FM and FIM ordering shift to lower temperatures with the introduction of Sn at the Zn sites in Mn$_{3}$ZnC \cite{teicher2019weyl}. Figures 1(c) and (d) show contour plots of isothermal magnetization $M(H)$ for MSGC and MSZC in the temperature range of 1.8 - 320 K. For MSGC, the $M(H)$ curves exhibit saturation up to $\sim$ 180 K, suggesting FM ordering. The value of saturated magnetic moment for MSGC is $\sim$ 0.24 $\mu_{\text{B}}$/Mn at 1.8 K. Similarly, for MSZC, the $M(H)$ curves saturate up to $\sim$ 200 K and the value of saturated magnetic moment is $\sim$ 0.32 $\mu_{\text{B}}$/Mn at 1.8 K (See Figure S4). The small fractional magnetic moments of the Mn atom for both compounds point to a strong itinerant character of the d-electrons.

\begin{table}[tb]
\text{Table 1. The fitting parameters derived from expression (1).}\
\centering
\resizebox{\columnwidth}{!}{ 
\begin{tabular}{c c c}
\hline
\textbf{Sample} & \textbf{MSGC} & \textbf{MSZC} \\
\hline
$\rho_{0}$ ($m\Omega \text{-cm}$) & 0.5 $\pm$ 0.00061 & 0.40 $\pm$ 0.00051 \\
$a$ (${10^{-6}}$$m\Omega \text{-cm} \, \text{K}^{-2}$) & 3.71 $\pm$ 0.34 & 1.04 $\pm$ 0.14 \\
$b$ (${10^{-5}}$m$\Omega \text{-cm} \, \text{K}^{-2/3}$) & 1.78 $\pm$ 0.85 & 4.16 $\pm$ 0.21 \\
$c$ (${10^{-5}}$m$\Omega \text{-cm} \, \text{K}^{-5/3}$) & 2.15 $\pm$ 0.41 & -- \\
$d$ (m$\Omega \text{-cm}$) & 0.452 $\pm$ 0.0021 & 1.16 $\pm$ 0.21 \\
$\theta_{D}$ (K) & 205.7 $\pm$ 4.04 & 172.7 $\pm$ 3.01 \\
\hline
\end{tabular}
}
\end{table}

\begin{figure*}
\includegraphics[width= 17 cm, height = 12 cm]{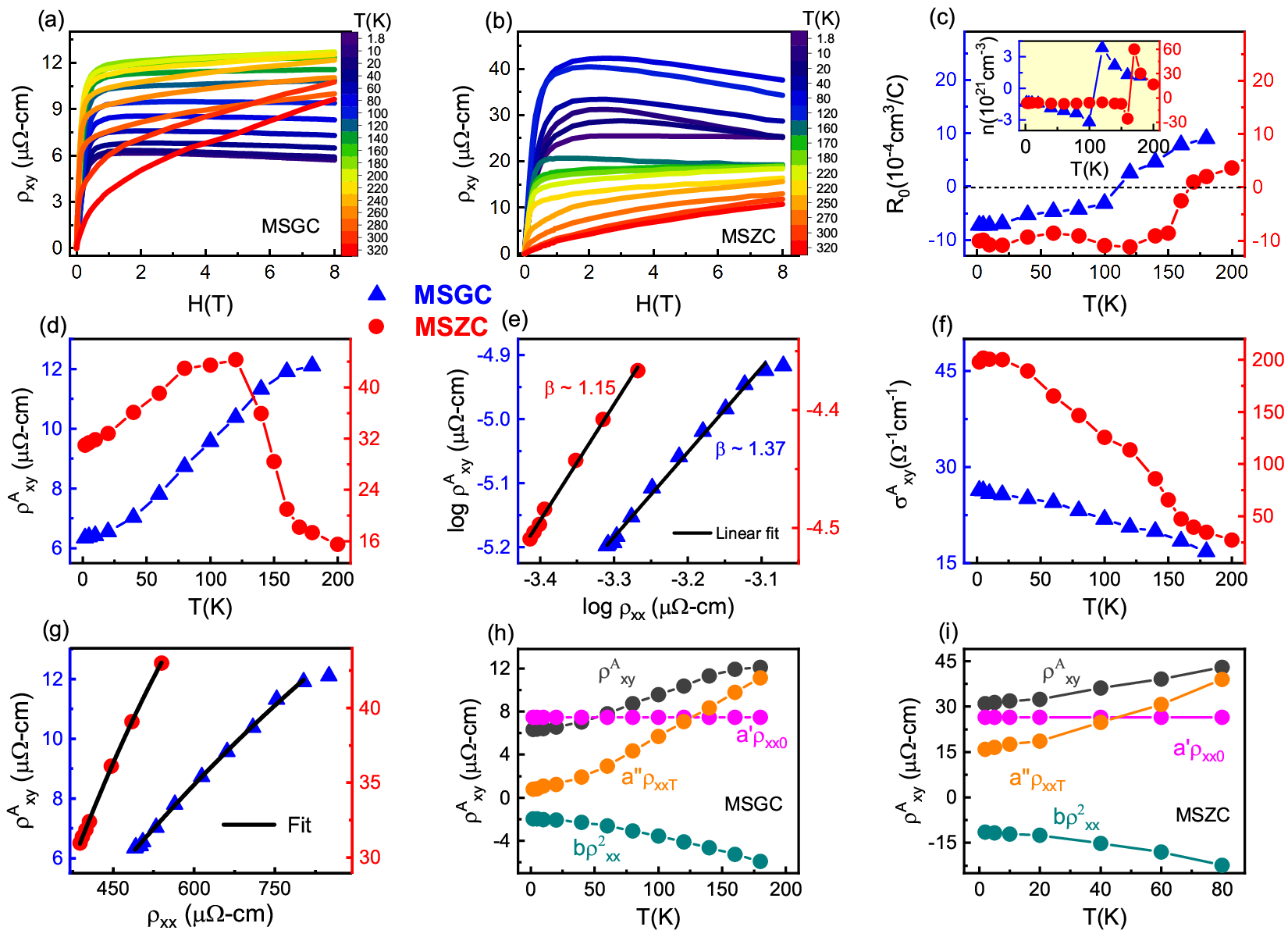}
\caption{(a) and (b) Field dependent Hall resistivity $\rho_{xy}$ of MSGC and MSZC at different temperatures. (c) Temperature-dependent normal Hall coefficient $R_{0}$. The inset shows the  carrier concentration $n$ as a function of temperature. (d) Temperature dependent anomalous Hall resistivity $\rho^A_{xy}$. (e) Plot between log $\rho^A_{xy}$ and $\rho_{xx}$. The solid black line indicate the fitting using the relation $\rho^A_{xy} \propto \rho^\beta_{xx}$. (f) Temperature-dependent anomalous Hall conductivity $\sigma^A_{xy}$. (g) Plot between $\rho^A_{xy}$ and $\rho_{xx}$, and the solid black line is the fitting result using $\rho^A_{xy} = a'\rho_{xx0} + a''\rho_{xxT} + b\rho^2_{xx}$. (h) and (i) The different contribution of $\rho^A_{xy}$ derived from the above expression for MSGC and MSZC, respectively.}
\label{fig:Figure1}
\end{figure*}

The longitudinal electrical resistivity ($\rho_{xx}$) of MSGC and MSZC, are shown in Fig. 1(a) and (b), respectively. The $\rho_{xx}(T)$ exhibits kink-like anomalies at $\sim$ 290 K for MSGC and $\sim$ 170 K for MSZC, related to the change in spin scattering across the magnetic transitions \cite{huang2022magnetoresistance,swanson1961electrical}. The $\rho_{xx}(T)$, is analyzed for different scattering mechanisms: electron-electron scattering, electron-magnon scattering and electron-phonon scattering. The $\rho_{xx}(T)$ data is fitted in the temperature range of 1.8 - 100 K for MSGC using the expression \cite{nascimento2018magnon,mukhopadhyay2021anomalous}
\begin{equation}
    \rho_{xx}(T) = \rho_0 + a T^{2} + b T^{3/2} + c T^{5/3} + \rho_{BG}
\end{equation}
where, $\rho_0$, $a T^{2}$, $b T^{3/2}$, $c T^{5/3}$ and $\rho_{BG}$ terms represent the residual resistivity, electron-electron scattering, electron-AFM magnon scattering, electron-FM magnon scattering and Bloch Gruneisen resistivity (electron-phonon scattering), respectively. The $\rho_{BG}$ is expressed as \cite{zhou2023structures}

\begin{equation}
 \rho_{BG} = d \left( \frac{T}{\theta_D} \right)^5 \int_0^{\frac{\theta_D}{T}} \frac{x^5}{(e^x - 1)(1 - e^{-x})} \, dx   
\end{equation}
where $\theta_D$ and $d$ represent the Debye temperature and constant related to electron-phonon strength, respectively. The expression (1) [solid cyan line in Fig. 1(a) and (b)] gives a good fit to the experimental data. It is to mention that For MSZC, $T^{5/3}$ term was excluded from expression due to the absence of FM-moments at low temperature in this compound. The obtained fitting parameters are shown in Table 1. It is evident that both compounds have significant contribution of electron-phonon and electron-magnon scattering.

Figure 2(a) and (b) show Hall resistivity ($\rho_{xy}$) as a function of field over a temperature range of 1.8 - 320 K. In general, the $\rho_{xy}$ of FM systems includes an additional contribution from the spontaneous magnetization, and can be expressed as \cite{bera2023anomalous,nagaosa2010anomalous,huang2023enhanced,roy2020anomalous,yu2025large}
\begin{equation}
    \rho_{xy}(H) = \rho_{0} + \rho^{A}_{xy} = R_{0}H + R_{S}M_{S}
\end{equation}
where $\rho_{0}$, $\rho^{A}_{xy}$, $R_{0}$, $R_{S}$ and $M_{S}$ represent the ordinary Hall resistivity, anomalous Hall resistivity, ordinary Hall coefficient, anomalous Hall coefficient and saturation magnetization, respectively. The $\rho_{xy}(H)$ is antisymmetrized using the expression $\rho_{xy}(H) = (\rho_{xy}(+H) + \rho_{xy} (-H))/2$ to remove the longitudinal components. The anomalous and ordinary Hall components are separated by linear fitting of $\rho_{xy}(H)$ in the saturation region, where the slope gives $R_0$ and the intercept gives $\rho_{xy}^A$. It is important to note that the estimation of $R_0$ and $\rho_{xy}^A$ from $\rho_{xy}(H)$ is possible only if the $M(H)$ curves saturating behavior \cite{bera2023anomalous,roy2020anomalous}. Therefore, we analyzed $\rho_{xy}(H)$ data up to 180 K for MSGC and up to 200 K for MSZC. The $R_0$ shown in Fig. 2(c) suggests the contribution from both electron and hole carriers, indicating a multi-band character for both compounds. The carrier density $n$ (inset of Fig. 2(c)), calculated using $R_0 = 1/ne$, shows similar temperature dependence for both compounds, with values of $\sim$ 1.38 $\times$$10^{22}$ cm$^{-3}$ for MSGC and $\sim$ 6.3$\times$$10^{21}$ cm$^{-3}$ for MSZC at 1.8 K. As shown in Fig. 2(d), $\rho^A_{xy}(T)$ for MSGC increases on increasing the temperature with a maximum value of $\sim$ 12 $\mu\Omega$-cm at 180 K. Whereas, for MSZC, $\rho^A_{xy}(T)$ shows a peak with the maximum value of $\sim$ 45 $\mu\Omega$-cm at 120 K. The magnitude of $\rho^A_{xy}(T)$ increases significantly with Ge and Zn doping, in comparison to Mn$_{3}$SnC \cite{huang2022magnetoresistance}.  
\begin{figure}
	\begin{center}
		\includegraphics[width=8.0cm]{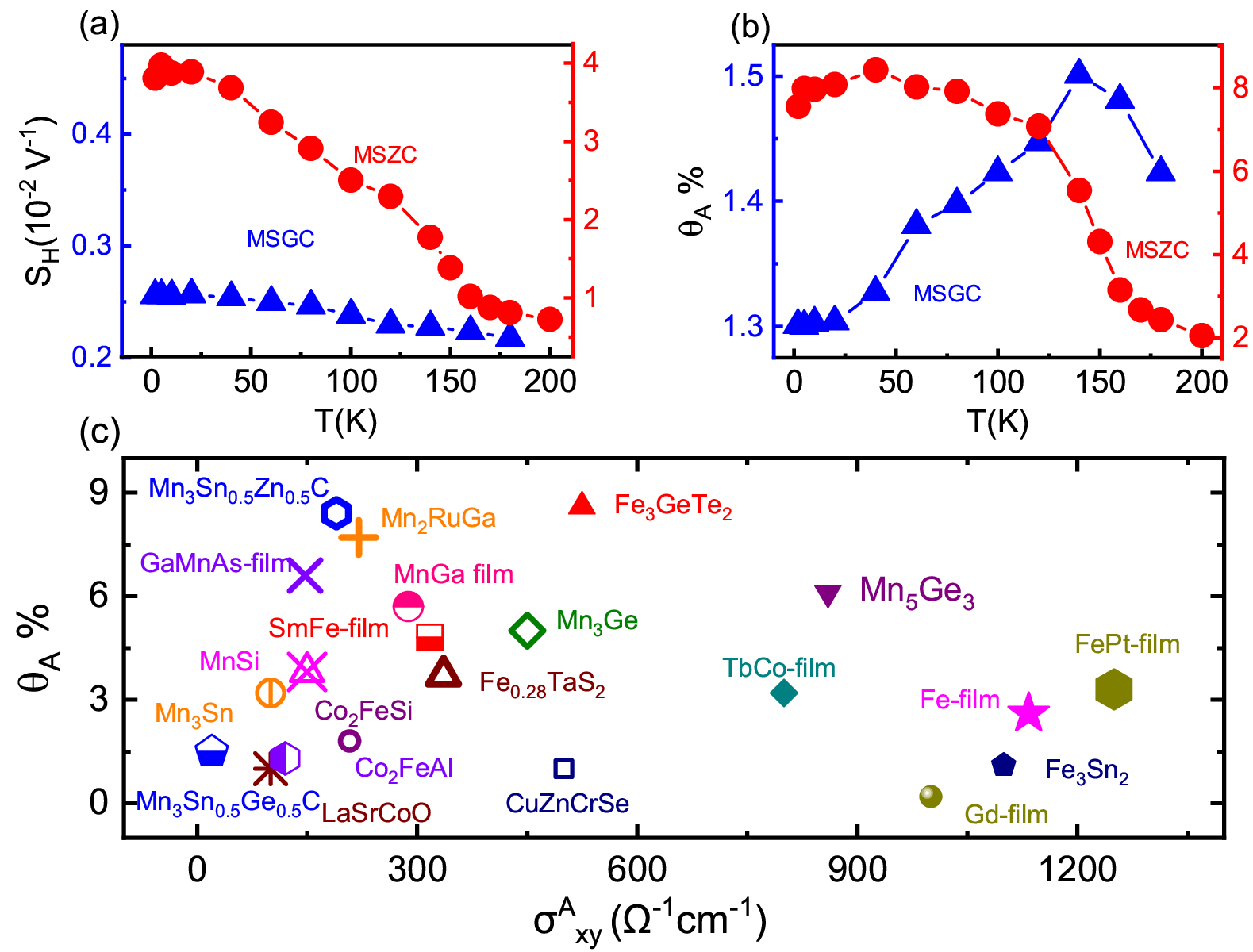}
		\caption{\label{Fig3} Plot of $\theta_{A}$ \% vs $\sigma^A_{xy}$ for present compounds and reported compounds in the literature. The references of the reported compounds are mentioned in the supplementary information.}
	\end{center}
\end{figure}

Generally, $\rho^A_{xy} \propto \rho^\beta_{xx}$, where the value of exponent $\beta$ = 1 and 2 corresponds to skew scattering and intrinsic Berry curvature or side jump, respectively \cite{bera2023anomalous,huang2023enhanced,roy2020anomalous}.
Therefore, We have plotted $\rho^A_{xy}$ and $\rho^A_{xx}$ on double logarithmic scale in Fig 2(e) to calculate the exponent $\beta$ using the expression $\rho^A_{xy} = \lambda\rho^\beta_{xx}$. where $\lambda$ is the SOC constant. The fitting range is restricted to 80 K in the MSZC, as $\rho^A_{xy}$ start to decrease beyond this temperature. This fitting range is similar to that employed for Fe$_{4}$GeTe$_{2}$ \cite{bera2023anomalous}, PrMn$_{2}$Ge$_{2}$ \cite{huang2023enhanced} and Co$_{3}$Sn$_{2}$S$_{2}$ \cite{liu2018giant}. The obtained values of $\beta$; 1.37 $\pm$ 0.087 (MSGC) and 1.15 $\pm$ 0.03 (MSZC), emphasis the presence of both skew scattering and intrinsic Berry curvature contributions in the AHE for both compounds.

Further, we examine the temperature dependence of AHC, which can be expressed by $\sigma^A_{xy} = \frac{\rho^A_{xy}}{(\rho^A_{xy})^2 + (\rho_{xx})^2}$ \cite{nagaosa2010anomalous,bera2023anomalous,roy2020anomalous,liu2018giant}. AHC decreases on increasing temperature for both compounds (Fig. 2(f)), with the maximum values of 26 $\Omega^{-1}$ cm$^{-1}$ and 200 $\Omega^{-1}$ cm$^{-1}$ at $T$ = 1.8 K for MSGC and MSZC, respectively. Although carbon non-stoichiometry can influence AHE and introduce a minor contribution to skew scattering, AHE in Mn$_{3}$SnC is mainly driven by the Berry curvature \cite{huang2022magnetoresistance}. The $\sigma^A_{xy}(T)$ reflects a dominant contribution of the skew scattering in the origin of the AHE of the Ge and Zn doped compounds. Due to the observation of combined contribution of skew scattering and Berry curvature contribution, we applied the scaling relation $\rho^A_{xy} = a\rho_{xx} + b\rho^2_{xx}$, where $a\rho_{xx}$ and $b\rho^2_{xx}$ represent the skew scattering and Berry curvature contribution, respectively \cite{bera2023anomalous,roy2020anomalous}. However, this scaling relation does not fit the experimental data well, which is possible if contribution from defects and phonon contribute separately \cite{luo2024magnetic, qi2021enhancement}. At low temperature $\rho_{xx} (T)$ is governed by electron-phonon and electron-magnon scattering, it is imperative to consider the contribution of these factors to skew scattering and $\rho^A_{xy}$. Several compounds, viz. MnGaGe \cite{luo2024magnetic}, CoFeMnSn \cite{xia2023structural} and PrCrGe$_{3}$ \cite{yu2025large} have been reported to exhibit electron-phonon contribution to skew scattering. As per Matthiessen rule for metals, $\rho_{xx}$ of metals consists a temperature independent residual resistivity resulting from impurity scattering, and the temperature-dependent resistivity, arising from the scattering of electrons with the excitations like phonons and magnons \cite{qi2021enhancement}. Here, the contributions from defects, phonons and magnons are treated on an equal basis $\rho^A_{xy}$, and the contribution of skew scattering ($\rho^{sk}_{xy}$) to $\rho^A_{xy}$ can be written as $\rho^{sk}_{xy} = a'\rho_{xx0} + a''\rho_{xxT}$, where $\rho_{xx0}$ represents the residual resistivity and $\rho_{xxT} = \rho_{xx} - \rho_{xx0}$. Tian $et$ $al.$ proposed a new scaling relation, known as Tian-YE-Jin (TYJ) model to extract the different contributions of AHE \cite{tian2009proper}. Therefore, we fitted $\rho^A_{xy}$ data using TYJ scaling model (Fig. 2(g)) up to 180 K for MSGC and up to 80 K for MSZC; 

\begin{equation}
    \rho^A_{xy} = a'\rho_{xx0} + a''\rho_{xxT} + b\rho^2_{xx}
\end{equation}
where $a'\rho_{xx0}$ and $a''\rho_{xxT}$ represent the contribution of defect induced scattering, electron-phonon and electron-magnon scattering, respectively. The $b\rho^2_{xx}$ term represents the combined contribution from the intrinsic Berry curvature and the side-jump mechanism. The contributions of different scattering mechanisms to $\rho^A_{xy}$ are plotted in Fig. 2(h) and (i) for MSGC and MSZC, respectively. These plots reveal that as the temperature rises, the electron-phonon and electron-magnon scattering contribution becomes more significant. Thus, AHE is governed by a combination of intrinsic Berry curvature and skew scattering, with additional contributions likely arising from defects, possibly associated with carbon deficiency in the compounds.

A second approach to identify the dominant contribution to the AHE involves examining the relationship between $\sigma^A_{xy}$ and $M_{S}$, and can be expressed as $S_{H} = R_{S}/\rho^2_{xx} = \sigma^A_{xy}/M_{S}$ (Fig 3(a)) \cite{bera2023anomalous,roy2020anomalous}. For intrinsic Berry curvature mechanism, $S_{H}$ is temperature-independent, but the strong temperature dependence in these compounds indicates that skew scattering has dominant contribution to the AHE of these doped compounds. Further, we calculate the anomalous Hall angle ($\theta_{A}$), which represents the relative contribution of the anomalous Hall current to the longitudinal current. This angle indicates the strength of a material for AHE, and is determined by the relation $\theta_{A} = \sigma^A_{xy}/\sigma_{xx}$ \cite{bera2023anomalous,roy2020anomalous,huang2023enhanced}. Temperature-dependent variation of $\theta_{A}(T)$ is plotted in Fig 3(b), and the maximum values of $\theta_{A}$ are found to be 1.5 \% and 8 \% for MSGC and MSZC, respectively. Temperature dependence of $\theta_{A}$ is also indicative of a skew scattering-dominated AHE in these compounds. We have shown the $\theta_{A}$ of these compounds  along with the previously reported results for other materials in Fig 3(c). The observed AHA in MSZC is greater than that for Mn$_{3}$Sn \cite{nakatsuji2015large} and Mn$_{3}$Ge \cite{nayak2016large}, and is comparable to Fe$_{3}$GeTe$_{2}$ \cite{bera2023anomalous}. MSGC has similar value of $\theta_{A}$ as Mn$_{3}$Sn.

In summary, we have explored magnetic and electronic transport properties of MSGC and MSZC. The MSGC exhibits concurrent AFM/FM state below $T_{C}$ $\sim$ 300 K, whereas, MSZC shows a FM transition at $T_{C}$ $\sim$ 240 K, followed by a FIM transition at $T_{N}$ $\sim$ 170 K. In the temperature range of 1.8 - 100 K, $\rho_{xx}(T)$ shows contributions from both electron-magnon and electron-phonon scattering. The doping of Ge and Zn atoms influences the magnetic transitions of Mn$_{3}$SnC and alters the key contributions to the origin of the AHE. Both intrinsic and extrinsic contributions are involved in the origin of the AHE. Electron-phonon and electron-magnon scattering plays a significant role in the skew-scattering mechanism. The present study offers an excellent platform for exploring the effect of electron-phonon and electron-magnon scattering on the AHE.

\section*{Supplementary Material}

See supplementary Material for the structural analysis, Hall resistivity and isothermal magnetization data of the studied compounds.

\section*{Acknowledgement}

We acknowledge Advanced Material Research Center (AMRC), IIT Mandi for the experimental facilities. SG acknowledges Ministry of Education, India for research fellowship. CSY acknowledges SERB-DST (India) for the CRG grant (CRG/2021/002743).

\section*{AUTHOR DECLARATIONS}
The authors declare no competing interests.
\section*{Author Contributions}
SG prepared the material and performed the experimental measurements. SG and CSY analyzed the data and wrote the manuscript. CSY supervised the overall project.
\section*{Data Availability}
The data that support the findings of this study are available from the corresponding authors upon reasonable request.

\nocite{*}
\bibliography{Mn3Sn0.5X0.5C}

\end{document}